# ALMA Memo 627

# Compilation of Technical Papers on ALMA Receivers

Tom Bakx[1] and John Conway[2], Chalmers University of Technology, Sweden

14th November 2024

**1. Introduction.** The Atacama Large Millimeter/submillimeter Array (ALMA) is the world's leading instrument for high-resolution imaging of the 0.3 to 10 mm wavelength sky. This interferometer exemplifies successful international collaboration even down to its individual components, including its telescope dishes and receivers produced in global partnerships. Over the past roughly ten years of ALMA operation, these receivers have been responsible for nearly ten thousand publications, as carefully monitored by the online ALMA Science Archive (see Figure 1). However, the citations of the relevant receiver papers have not managed to grow together with their use in the astronomical community, which points out an information gap among the astronomical community. To overcome this gap, this memo provides a comprehensive list of receiver references, which was created in direct contact with the receiver groups.

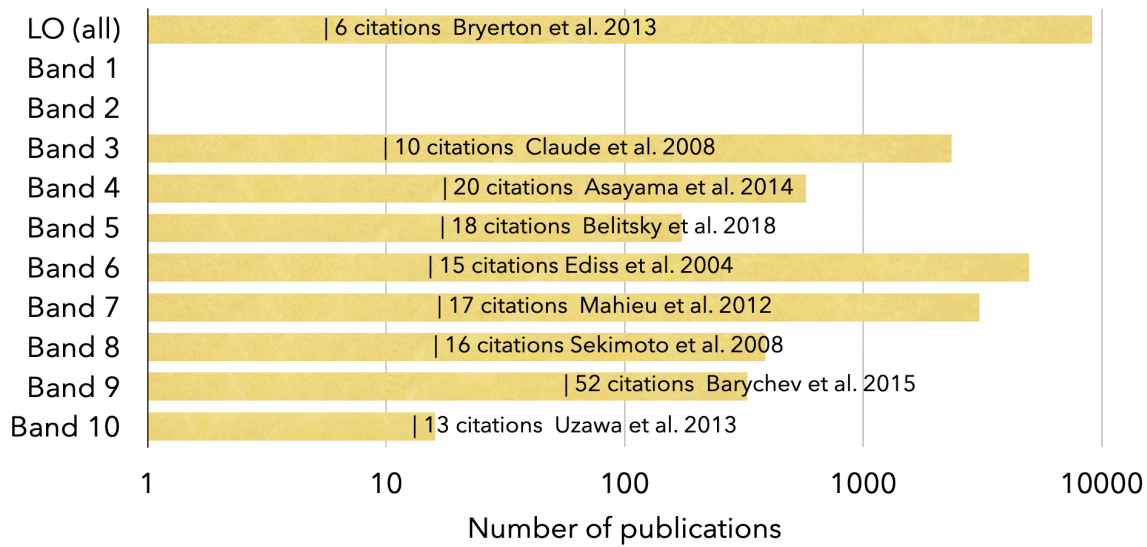

Figure 1: The number of publications (as of May 2024) associated with each band according to the ALMA science archive after accounting for duplicates and multi-band observations. The number of citations of the receiver description paper is overlaid, showing a discrepancy between the scientific usage of the ALMA receivers and the academic recognition of the instrumentalists who have built these receivers caused by an information gap.

## 2. Compiled list of Technical Papers on ALMA Receivers

In this memo, we aim to facilitate access to a list of technical references (see Table 1) to address the citation disparity. This table was compiled through a direct discussion with the relevant receiver groups. The acknowledgement of the usage of ALMA receivers can already be deduced from the online ALMA science archive, which provides a detailed overview of the relevant receiver usage statistics across ALMA's wide scientific goals. An increase in the number of citations of the relevant ALMA receivers as well as the central correlator will further facilitate community-wide recognition and, for some instrumentation groups, increase access to academic funding sources, which in turn can help sustain future innovation in receiver technology providing great benefit to the whole ALMA community. The authors note that the receiver citation information is already (partially) available in Table 4.2 (page 31 of the current version) of the ALMA Technical Handbook at [https://almascience.eso.org/documents-and-tools/cycle11/alma-technical-handbook](https://almascience.eso.org/documents-and-tools/cycle11/alma-technical-handbook). This memo does not reflect official ALMA policy and aims to complement this official ALMA documentation.

The authors identified the origins of the citation disparity as due to an information gap, both through their direct experience, as well as through extensive discussions within the astronomical community. In part, this gap is caused by the use of different journals across the instrument builder (e.g., IEEE) and observing (e.g., A&A, ApJ) communities. A first step to addressing the citation disparity would thus be the compilation of a list of references together with the instrument building community.

Table 1 lists for each receiver band the compiled references, and has been produced through direct communication with each relevant receiver group. A reference to the local oscillator development common to all receiver bands (excluding future Band 2) is also included, as its design and subsequent performance is central to the functioning of ALMA. For Bands 1 and 2, the final papers describing these receivers are in preparation with likely completion within approximately 6 months; in the meantime, the relevant receiver groups have confirmed that the references given in Table 1 should be used. The ALMA Helpdesk can further provide information for the ALMA receiver papers, in particular regarding the Band 1 and 2 receivers. For some receiver bands, multiple references together give the full technical description of the receiver; these additional references are given in the Appendix Table 2.

Table 1: Compiled paper references for ALMA receivers (Cit. indicates the total number of citations to the listed paper in both the astronomy and instrumentation literature, and Pub. gives the number of astronomy papers that have made use of that receiver as of May 2024).

| Band | Reference | Cit. | Pub. | link |
|---|---|---|---|---|
| all: Local Oscillator | Bryerton et al. 2013 | 6 | 9138 | http://ieeexplore.ieee.org/stamp/stamp.jsp?tp=&arnumber=6697622 |
| 1 | Huang et al. 2022 * | 1 | 0 | https://ui.adsabs.harvard.edu/abs/2022SPIE12190E..0KH/abstract |
| 2 | Yagoubov et al. 2020 * | 24 | 0 | https://ui.adsabs.harvard.edu/abs/2020A%26A...634A..46Y/abstract |
| 3 | Claude et al. 2008 | 10 | 2342 | https://ui.adsabs.harvard.edu/abs/2008SPIE.7020E..1BC/abstract |
| 3+6 | Kerr et al. 2014 | 31 | 6422 | https://ui.adsabs.harvard.edu/abs/2014ITTST...4..201K/abstract |
| 4 | Asayama et al. 2014 | 20 | 575 | https://ui.adsabs.harvard.edu/abs/2014PASJ...66...57A/abstract |
| 5 | Belitsky et al. 2018 | 18 | 173 | https://ui.adsabs.harvard.edu/abs/2018A%26A...611A..98B/abstract |
| 6 | Ediss et al. 2004 | 15 | 4953 | https://ui.adsabs.harvard.edu/abs/2004stt..conf..181E/abstract |
| 6 | Kerr et al. 2004 | 9 | 4953 | http://www.nrao.edu/meetings/isstt/papers/2004/2004055061.pdf |
| 7 | Mahieu et al. 2012 | 17 | 3064 | https://ui.adsabs.harvard.edu/abs/2012ITTST...2...29M/abstract |
| 8 | Sekimoto et al. 2008 | 16 | 388 | https://ui.adsabs.harvard.edu/abs/2008stt..conf..253S/abstract |
| 9 | Baryshev et al. 2015 | 52 | 327 | https://ui.adsabs.harvard.edu/abs/2015A%26A...577A.129B/abstract |
| 10 | Uzawa et al. 2013 | 13 | 16 | https://ui.adsabs.harvard.edu/abs/2013PhyC..494..189U/abstract |

Note * - The proposed papers to cite for Band 1 and 2 are preliminary pending the publication of final instrument papers that are currently in preparation, (expected within 6 to 8 months).

**Appendix - Additional references**

This appendix gives additional receiver references (Table 2) of the current ALMA system.

Table 2: Additional receiver references

| Paper focus | Reference | Cit. | link |
| --- | --- | --- | --- |
| General North-American receiver papers | Effland et al. 2013 | 0 | http://ieeexplore.ieee.org/stamp/stamp.jsp?tp=&arnumber=6697565 |
| North-American SIS mixers | Kerr et al. 2013 | 4 | http://ieeexplore.ieee.org/stamp/stamp.jsp?tp=&arnumber=6697439 |
| Additional Band 3 receiver paper | Pan et al. 2004 | 9 | http://www.nrao.edu/meetings/isstt/papers/2004/2004062069.pdf |
| Additional Band 3 receiver paper | Claude et al. 2006 | 10 | https://ui.adsabs.harvard.edu/abs/2006stt..conf..154C/abstract |
| Additional Band 3 receiver paper | Claude et al. 2005 | 6 | http://ieeexplore.ieee.org/stamp/stamp.jsp?tp=&arnumber=1572585 |
| Additional Band 3 receiver paper | Dindo et al. 2005 | 5 | http://ieeexplore.ieee.org/stamp/stamp.jsp?tp=&arnumber=1572586 |
| Additional Band 3 Side-band receiver paper | Chin et al. 2004 | 9 | https://link.springer.com/article/10.1023/B:IJIM.0000020748.79086.e9 |
| Additional Band 5 receiver | Billade et al. 2012 | 35 | https://ui.adsabs.harvard.edu/abs/2012ITTST...2..208B/abstract |
| Additional Band 7 receiver paper | Maier et al. 2005 | 15 | https://ui.adsabs.harvard.edu/abs/2005stt..conf..428M/abstract |
| Additional Band 7 receiver paper | Mahieu et al. 2005 | 10 | https://ui.adsabs.harvard.edu/abs/2005stt..conf...99M/abstract |
| Additional Band 8 receiver paper | Shan et al. 2005 | 15 | https://ui.adsabs.harvard.edu/abs/2005ITAS...15..503S/abstract |
| Additional Band 8 receiver paper | Sekimoto et al. 2009 | 2 | https://ui.adsabs.harvard.edu/abs/2009stt..conf....6S/abstract |
| Additional Band 8 receiver paper | Tamura et al. 2014 | 5 | http://ieeexplore.ieee.org/stamp/stamp.jsp?tp=&arnumber=6945347 |
| Additional Band 9 receiver paper | Baryshev et al. 2007 | 4 | https://ui.adsabs.harvard.edu/abs/2007stt..conf..164B/abstract |
| Additional Band 9 receiver paper | Mena et al. 2008 | 3 | https://ui.adsabs.harvard.edu/abs/2008stt..conf...90M/abstract |
| Additional Band 9 receiver paper | Baryshev et al. 2008 | 1 | https://ui.adsabs.harvard.edu/abs/2008stt..conf..258B/abstract |
| Additional Band 9 receiver paper | Hesper et al. 2018 | - | www.nrao.edu/meetings/isstt/papers/2018/2018098103.pdf |
| Additional Band 10 paper (summary of production) | Gonzalez et al. 2014 | 6 | https://ui.adsabs.harvard.edu/abs/2014SPIE.9153E..0NG/abstract |
| Additional Band 10 receiver paper | Uzawa et al 2009 | 2 | https://ui.adsabs.harvard.edu/abs/2009stt..conf...12U/abstract |
| Additional Band 10 receiver paper | Fujii et al. 2011 | 13 | https://ui.adsabs.harvard.edu/abs/2011ITAS...21..606F/abstract |
| Additional Band 10 receiver paper | Fujii et al. 2013 | 26 | http://ieeexplore.ieee.org/stamp/stamp.jsp?tp=&arnumber=6423869 |